\documentclass[pre,twocolumn,floatfix,showpacs,amsmath,amssymb]{revtex4}
\usepackage{graphicx}
\begin{document}
\title{The Decay of Magnetohydrodynamic Turbulence from Power-Law Initial 
Conditions}
\author{Chirag Kalelkar}
\email{kalelkar@physics.iisc.ernet.in}
\author{Rahul Pandit}
\email{rahul@physics.iisc.ernet.in}
\altaffiliation[\\ Also at ]{the Jawaharlal Nehru Centre for Advanced 
Scientific Research, Bangalore, India.}
\affiliation{Centre for Condensed Matter Theory,
Department of Physics, Indian Institute of Science, Bangalore 560012, India.}
\begin{abstract}
We derive relations for the decay of the kinetic and magnetic energies and the 
growth of the Taylor and integral scales in unforced, 
incompressible, homogeneous and isotropic three-dimensional 
magnetohydrodynamic (3DMHD) turbulence with power-law initial energy 
spectra. We also derive bounds for the decay 
of the cross- and magnetic helicities. We then present results from systematic 
numerical studies of such decay both within the context of an MHD shell 
model and direct numerical simulations (DNS) of 3DMHD. We show explicitly that 
our results about the power-law decay of the energies hold for times $t<t_*$, 
where $t_*$ is the time at which the integral scales become comparable to the 
system size. For $t<t_*$, our numerical results are consistent with those 
predicted by the principle of `permanence of large eddies'. 
\end{abstract}
\maketitle
\section{Introduction}
The decay of homogeneous and isotropic, \textit{fluid} turbulence has been the 
subject of extensive theoretical\cite{Decayth,Monin} and 
experimental\cite{Corrsin,Helium,Uberoi} studies. These include wind-tunnel 
experiments downstream of a grid\cite{Corrsin} and flow behind a towed grid
 in a stationary channel filled with Helium II\cite{Helium}. However, 
unambiguous statements can still not be made about the dependence of the decay 
of fluid turbulence on the initial conditions. In experiments, these can be 
changed to some extent by changing the grid geometry and choice of virtual 
origin\cite{Uberoi} but caution is advised in interpreting the results since 
our ability to choose initial conditions precisely is severely restricted. 
In an earlier study\cite{George}, the energy-decay 
exponents were predicted to be initial-condition dependent, but the precise 
nature of this dependence was not elucidated clearly. Direct numerical 
simulations (DNS) 
can control initial conditions, but, to the best of our knowledge, have 
investigated a very limited class of initial conditions. A recent 
exception\cite{Ditlevsen} is a study of the decay of fluid 
turbulence for the set of initial conditions with a power-law initial 
energy spectrum $E^0(k)\sim k^q$, where the superscript $0$ denotes the choice  
of virtual origin of time $t=t_0$, $k$ is the magnitude of the wave vector, and 
the exponent $q$ distinguishes different initial conditions in this set. 

Results, both experimental and numerical, for the decay of {\it 
magnetohydrodynamic} (MHD) turbulence are even more scarce than their 
fluid-turbulence analogs. 
Different power-law decays have been suggested in DNS\cite{Biskamp} studies, 
but the sensitivity of these decays to the initial conditions or evolution of 
the relevant length scale has not been investigated in any detail. 

In this paper, we initiate such an investigation 
and obtain the following interesting results: 
We show first how to generalise the results of Ref. \cite{Ditlevsen,Olesen} 
and derive, 
for unforced, incompressible, homogeneous and isotropic\cite{Isotropic} 
three-dimensional MHD (3DMHD) turbulence, expressions for the decay of the 
kinetic and magnetic energies and the growth of the Taylor and integral 
scales, when we start with power-law 
initial conditions $E^0_{\textbf a}(k)\sim k^q$ ($q>-1$), 
where the subscript {\textbf a} is {\textbf v} for the kinetic energy and 
{\textbf b} for the magnetic energy. Such initial conditions are of interest 
in the astrophysical context of the decay of power-law `primordial' energy 
spectra\cite{Olesen}. We also derive bounds for the decay of the cross- and 
magnetic helicity.We then show by systematic numerical studies of a shell 
model for MHD turbulence\cite{Frick,Basu} and an $80^3$ pseudospectral DNS of 
the 3DMHD equations that, given power-law 
initial conditions, the kinetic and magnetic energies and the Taylor and 
integral scales follow the decay expressions mentioned above within a regime 
governed by the temporal evolution of the integral scales. 

The unforced MHD equations are 
\begin{eqnarray}
&&\frac{\partial{\bf v}}{\partial{t}}+({\bf v\cdot\bigtriangledown}){\bf v}=
-\frac{\bigtriangledown p^*}{\rho}+\frac{({\bf b}\cdot
\bigtriangledown){\bf b}}{4\pi\rho}+\nu{\bigtriangledown^2\bf v},\nonumber\\
&&{\partial{\bf b}\over\partial t}=\bigtriangledown\times({\bf v}\times
{\bf b})+\eta\bigtriangledown^2{\bf b},
\label{mhd}
\end{eqnarray}
where $\nu$ is the kinematic viscosity, $\eta$ is the magnetic viscosity, 
$p^*=p+b^2/8\pi$ is the effective pressure, and $p$ is the pressure. We 
enforce the incompressibility condition $\bigtriangledown\cdot{\textbf v}=0$ 
and eliminate $p^*$ in the usual manner.
The invariants of inviscid, unforced 3DMHD are the total energy 
($E_T=E_{\textbf v}+E_{\textbf b}\equiv\frac{1}{2}\int
(v^2+b^2)\hspace{0.2cm}dV$), the cross-helicity ($H_C\equiv\int{\bf v\cdot 
b}\hspace{0.2cm}dV$), and the magnetic helicity 
($H_M\equiv\int{\bf A\cdot b}\hspace{0.2cm}dV$, where {\bf A} is the vector 
potential). We restrict ourselves to the case with $|H_M^0|$ close to 
zero and exclude any helical contributions\cite{Biskamp} to the 
decay process due to the inverse cascade of $H_M$. Most of our results are 
obtained for initial equipartition 
of energy, namely, $E_{\bf v}^0=E_{\bf b}^0$, but we also present a few results 
for $E_{\bf v}^0\neq E_{\bf b}^0$. 
\section{Theoretical Results}
We observe that Eqs. (\ref{mhd}) are invariant under the rescaling
\begin{eqnarray}
&&{\textbf x}\rightarrow l{\textbf x},t\rightarrow l^{1-h}t,{\textbf v}
\rightarrow l^h{\textbf v},{\textbf b}\rightarrow l^h{\textbf b},\nonumber\\
&&\nu\rightarrow l^{1+h}\nu,\eta\rightarrow l^{1+h}\eta,
\label{scale}
\end{eqnarray}
where $l$ is an arbitrary scale factor and $h<0$ a scaling exponent.
The local kinetic- and magnetic-energy densities in three dimensions are given 
by 
\begin{eqnarray}
&&{\mathcal E}_\textbf{a}(k,t,\nu,\eta,K,2\pi/L)=\nonumber\\
&&\frac{\Omega k^2}{(2\pi)^3L^3}\int_{2\pi/K}^Ld^3x
d^3y\hspace{0.2cm}e^{i{\bf k\cdot(x-y)}}\langle{\bf a}({\bf x},t)
\cdot{\bf a}({\bf y},t)\rangle,\hspace{0.7cm}
\label{define}
\end{eqnarray}
where $L$ and $2\pi/K$ are the large- and small-distance cutoffs, the 
subscript {\bf a} is {\bf v} for the kinetic energy density and {\bf b} for 
the magnetic energy density and $\Omega$ is the solid angle. Angular brackets 
denote an average over the initial conditions being considered (see below). 
Thus the kinetic and magnetic energies in this range of length scales are 
\begin{eqnarray}
E_{\textbf a}(t,\nu,\eta,2\pi/K,L)=\int_{2\pi/L}^K dk\hspace{0.2cm}
{\mathcal E}_{\textbf a}(k,t,\nu,\eta,K,2\pi/L).
\label{toten}
\end{eqnarray}
By combining Eqs.~(\ref{scale}) and (\ref{define}) we get 
\begin{eqnarray}
&&{\mathcal E}_{\textbf a}(k/l,l^{1-h}t,l^{1+h}\nu,l^{1+h}\eta,K/l,2\pi/Ll)=
 \nonumber\\
&&l^{1+2h}\hspace{0.2cm}{\mathcal E}_{\textbf a}(k,t,\nu,\eta,K,2\pi/L).
\label{changed}
\end{eqnarray}
Henceforth the dependence of ${\mathcal E}_{\textbf a}$ on $K$ and $2\pi/L$ 
will be suppressed for notational convenience. It is useful to define the 
functions
\begin{eqnarray}
\Phi_{\textbf a}(k,t,\nu,\eta)\equiv k^{1+2h}{\mathcal E}_{\textbf a}
(k,t,\nu,\eta)
\label{changed1}
\end{eqnarray}
in terms of which Eq. (\ref{changed}) becomes
\begin{eqnarray}
\Phi_{\textbf a}(k/l,l^{1-h}t,l^{1+h}\nu,l^{1+h}\eta)=\Phi_{\textbf a}
(k,t,\nu,\eta);
\label{prev}
\end{eqnarray}
when differentiated with respect to $l$, this yields, in the limit 
$l\rightarrow1$,
\begin{eqnarray}
&&-\frac{\partial\Phi_{\textbf a}}{\partial\ln k}+(1-h)\frac{\partial\Phi
_{\textbf a}}{\partial\ln t}
+(1+h)\frac{\partial\Phi_{\textbf a}}{\partial\ln\nu}+\nonumber\\
&&(1+h)\frac{\partial\Phi_{\textbf a}}{\partial\ln\eta}=0.
\end{eqnarray}
The general solution of the above equation is
\begin{eqnarray}
&&\Phi_{\textbf a}(k,t,\nu,\eta)=F_{\bf a}((1-h)\ln k+\ln t,(1+h)
\ln t+\nonumber\\
&&(-1+h)\ln\nu,(1+h)\ln t+(-1+h)\ln\eta).
\end{eqnarray}
This solution corresponds to 
\begin{eqnarray}
&&{\mathcal E}_{\textbf a}(k,t,\nu,\eta)=
k^q\hspace{0.2cm}\Phi_{\textbf a}(k^{(3+q)/2}t,t^{(1-q)/2}
\nu^{-(3+q)/2},\nonumber\\
&&t^{(1-q)/2}\eta^{-(3+q)/2}),
\label{main}
\end{eqnarray}
where $q=-1-2h$ ($q>-1$)\cite{Convergence}.\\
For $q=1$, Eqs. (\ref{toten}) and (\ref{main}) lead to especially simple 
forms for the temporal decay of the kinetic- and magnetic-energies. We are 
interested in the limits $2\pi/L\rightarrow0$ and $K\rightarrow\infty$, of 
relevance to high-Reynolds number turbulence, so
\begin{eqnarray}
E_{\textbf a}(t)=
\frac{1}{2t}~\int_0^\infty dy ~\Phi_{\textbf a}(y,1/\nu^2,1/\eta^2).
\label{eqb}
\end{eqnarray}
Thus both $E_{\textbf v}$ and $E_{\textbf b}$ decay as $1/t$, which 
generalizes the result of Ref. \cite{Ditlevsen} to the case of decaying 
MHD turbulence. For all 
$q\neq1$, the simple power-law dependence of $E_{\bf a}(t)$ on $t$ does not 
follow. The fluid and magnetic Taylor scales, 
$\lambda_{\textbf a}(t,\nu,\eta,2\pi/K,L)
\equiv[\int_{2\pi/L}^K dk\hspace{0.2cm}{\mathcal E}_{\textbf a}
(k,t,\nu,\eta)/\int_{2\pi/L}^K dk\hspace{0.2cm}k^2\mathcal E_{\textbf a}(k,
t,\nu,\eta)]^{1/2}$, also show a simple power-law dependence on $t$ only for 
$q=1$. In the limits $2\pi/L\rightarrow0$, $K\rightarrow\infty$,
\begin{eqnarray}
\lambda_{\textbf a}(t)=t^{1/2}\left[\frac{\int_0^\infty dy\hspace{0.1cm}
\Phi_{\textbf a}(y,1/\nu^2,1/\eta^2)}{\int_0^\infty dy\hspace{0.1cm}
\Phi_{\textbf a}(y,1/\nu^2,1/\eta^2)y}
\right]^{1/2}.
\label{taylor}
\end{eqnarray}
Similarly, the integral scales $L_{\bf a}(t,\nu,\eta,2\pi/K,L)\equiv
[\int_{2\pi/L}^K dk\hspace{0.2cm}{\mathcal E}_{\textbf a}(k,t,\nu,\eta)/k]
/\int_{2\pi/L}^K dk\hspace{0.2cm}\mathcal E_{\textbf a}(k,t,\nu,\eta)$ (where 
the subscript {\bf a} is {\bf v} for the fluid integral scale and {\bf b} for 
the magnetic integral scale), also grow as a power of $t$ with exponent equal 
to $0.5$ for $q=1$ in the limits $2\pi/L\rightarrow0$, $K\rightarrow\infty$. 

A consequence of the positive-definiteness of the energy spectra is that 
the helicity spectra satisfy realizability constraints, $|{\mathcal H_C}(k,t,
\nu,\eta)|\leq[{\mathcal E}_{\textbf v}(k,t,\nu,\eta){\mathcal E}_{\textbf b}
(k,t,\nu,\eta)]^{1/2}$, and, $|{\mathcal H}_M(k,t,\nu,\eta)|\leq{\mathcal E}_
{\textbf b}(k,t,\nu,\eta)/k$, where ${\mathcal H}_C(k,t,\nu,\eta)$ and 
${\mathcal H}_M(k,t,\nu,\eta)$ are the cross- and magnetic helicity densities 
defined like the energy densities (Eqs. (\ref{define})). 
Hence, in the limits $2\pi/L\rightarrow0$, $K\rightarrow\infty$ (with variables 
other than $k$ and $t$ suppressed for notational convenience), the 
cross-helicity (for $q=1$) satisfies
\begin{eqnarray}
|H_C(t)|\leq\int_0^\infty dk\hspace{0.1cm}|\mathcal H_C(k,t)|\nonumber\\
\leq\frac{1}{2t}\int_0^\infty dy\hspace{0.1cm}
[\Phi_{\textbf v}(y)\Phi_{\textbf b}(y)]^{1/2}
\label{hc}
\end{eqnarray}
and the magnetic helicity (for $q=1$) satisfies
\begin{eqnarray}
|H_M(t)|\leq\int_0^\infty dk\hspace{0.1cm}|\mathcal H_M(k,t)|\nonumber\\
\leq\frac{1}{2\sqrt{t}}\int_0^\infty dy\hspace{0.1cm}
\frac{\Phi_{\textbf b}(y)}{\sqrt{y}}.
\label{hm}
\end{eqnarray}
The results obtained above apply for times $t<t_*$, where $t_*$ is the 
crossover time at which $L_{\bf a}(t)$ becomes equal to the size of the system 
(or the linear size of the simulation box in a DNS). For $t>t_*$, finite-size 
effects, that might well be nonuniversal, modify the decay of $E_{\bf a}(t)$.
\section{Numerical Results}
We now report on the numerical studies we have carried out to check the results 
given above. In most of our runs, we concentrate on the region $0<t<t_*$. 
We use double-precision arithmetic, 
but have checked in representative cases that our results are unaffected if 
we use quadruple-precision arithmetic.

Shell models comprise a set of ordinary differential equations 
containing suitable non-linear coupling terms that respect the analogs 
of the conservation laws of the 3DMHD equations in the inviscid, 
unforced limit. 
They exhibit an energy cascade in the presence of viscosities. 
We confine our study to a shell model proposed in \cite{Frick,Basu}  
that enforces all the ideal 3DMHD invariants in the inviscid, unforced case, 
that reduces to the well-known GOY\cite{GOY} model for fluid turbulence when 
magnetic terms are suppressed, has no adjustable parameters apart from the 
fluid and magnetic Reynolds numbers, and exhibits the same multiscaling (within 
error bars) as the 3DMHD equations.
 
The unforced shell-model equations are\cite{Erratum}
\begin{eqnarray}
\frac{dz_n^{\pm}}{dt}=ic_n^{\pm}-\nu_+k_n^2z_n
^{\pm}-\nu_-k_n^2z_n^{\mp},
\label{shell}
\end{eqnarray}
with the complex, scalar
Els\"asser variables $z_n ^{\pm}\equiv(v_n\pm b_n)$, $i=\sqrt{-1}$, 
and the discrete wavenumbers $k_n=k_02^n$ ($k_0$ sets the scale for 
wave-numbers), for shell index $n$ ($n=1\ldots N$, for $N$ shells) with        
\begin{eqnarray}
c_n^{\pm}&=&[a_1k_nz_{n+1}^{\mp}z_{n+2}^{\pm}
+a_2k_nz_{n+1}^{\pm}z_{n+2}^{\mp}\nonumber\\
&+&a_3k_{n-1}z_{n-1}^{\mp}z_{n+1}^{\pm}
+a_4k_{n-1}z_{n-1}^{\pm}z_{n+1}^{\mp}\nonumber\\
&+&a_5k_{n-2}z_{n-1}^{\pm}z_{n-2}
^{\mp}+a_6k_{n-2}z_{n-1}^{\mp}z_{n-2}^{\pm}]^*\nonumber.
\end{eqnarray}
Here $a_1=7/12$, $a_2=5/12$, $a_3=-1/12$, $a_4=-5/12$, $a_5=-7/12$, $a_6=1/12$ 
and $\nu_{\pm}=\nu\pm\eta$.
Shell-model analogs of the total energy ($E_T^s=E_v^s+E_b
^s\equiv(1/2)\sum_n (|v_n|^2 + |b_n|^2)$), the cross
helicity ($H_C^s\equiv(1/2)\sum_n(v_nb_n^*+v_n^*b_n)$), and the
 magnetic helicity ($H_M^s\equiv\sum_n(-1)^n|b_n|^2/k_n$) are conserved if 
$\nu_{\pm}=0$. Here and in the following, the superscript $s$ stands for shell 
model. We solve Eqs.~(\ref{shell}) numerically by an Adams-Bashforth 
scheme\cite{Pisarenko} (step size $\delta t^s=10^{-2}$) and use $N=22$ shells 
with $k_0=1/16$ and $\nu=\eta=10^{-4}$. 

In decaying turbulence, it is convenient to measure time in units of the 
initial large eddy-turnover times. For our shell model these are 
$\tau_a\equiv1/(a_{rms}^{0,s}k_1)$ with $a_{rms}^{0,s}
\equiv[\langle\sum_n|a_n^0|^2\rangle]^{1/2}$, the root-mean-square values of 
the initial velocities and 
magnetic fields. Since $\tau_a$ are calculated at the 
start of the simulation runs, when the velocity and magnetic fields differ 
only in phase (see below), here $\tau_v=\tau_b$ equal 
$18.7$ for $q=1$ and $7.8$ for $q=0$.
Our runs are reported in terms of $\tau^s\equiv t/\tau_a$
($t$ is the product of the number of iterations and 
$\delta t^s$), and are ensemble averaged over $100$ independent initial 
conditions with varying phases. We define 
$Re_v^{0,s}\equiv v^{0,s}_{rms}/(k_1\nu)$ and $Re_b^{0,s}\equiv b^{0,s}_{rms}/
(k_1\eta)$ to be the values of the initial fluid and 
magnetic Reynolds numbers (here $Re_v^{0,s}$ and $Re_b^{0,s}$ 
equal $34246$ for $q=1$ and $81920$ for $q=0$). 
The initial velocity and magnetic fields are taken to be  
$v_n^0={k_n}^{(1+q)/2}e^{i\theta_n}$ and $b_n^0={k_n}^{(1+q)/2}e^{i\phi_n},$ 
with $\theta_n$ and $\phi_n$ being independent random variables distributed 
uniformly between $0$ and $2\pi$ (our shell-model energy densities being 
defined as ${\cal E}_a^s\equiv \langle|a_n|^2/k_n\rangle$). In all our runs, 
$|H_M^{0,s}|/H_M^{max,s}\lesssim10^{-6}$, where $H_M^{max,s}
\simeq E_T^{0,s}/k_1$.  

\begin{figure}
\includegraphics[height=2in]{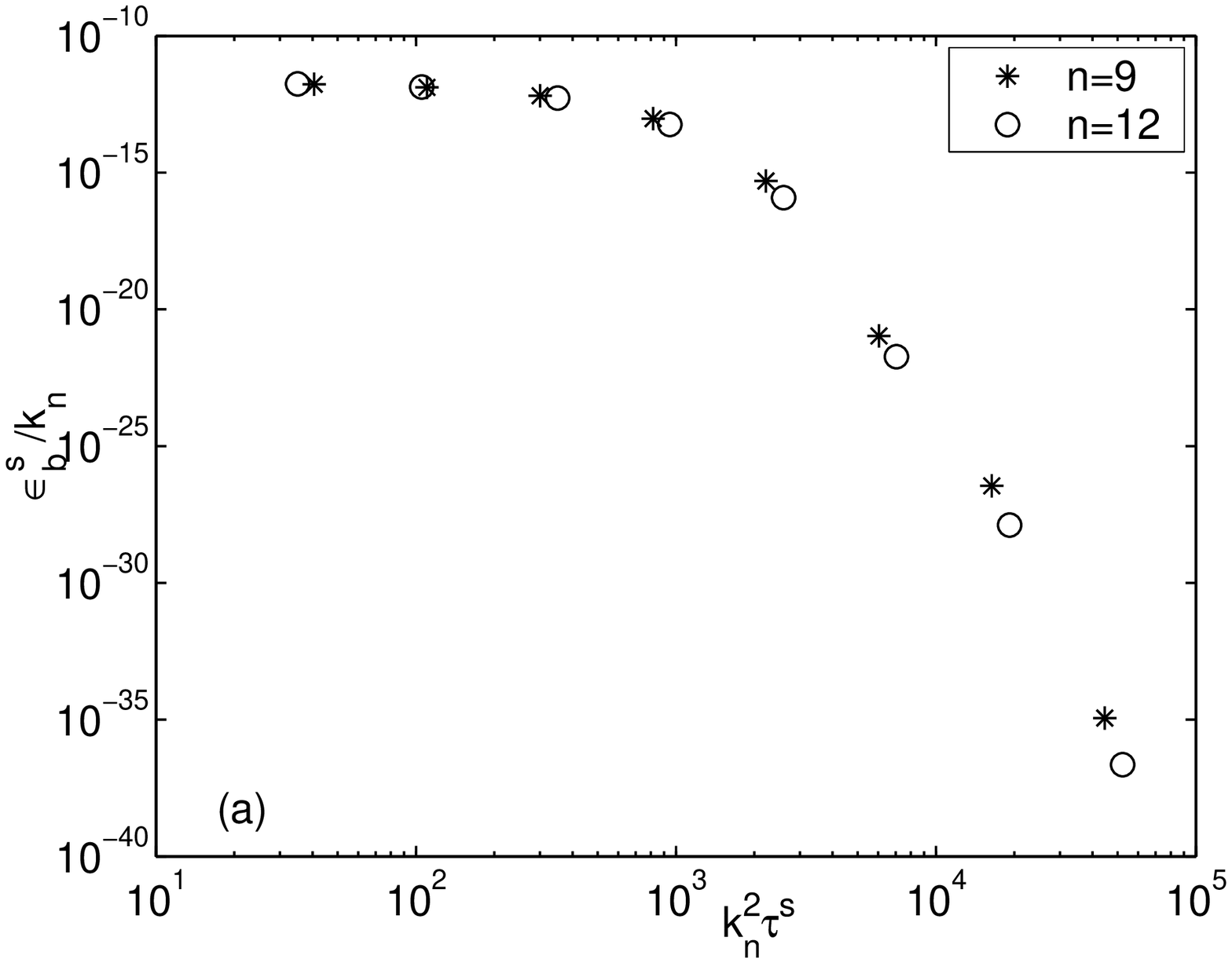}
\includegraphics[height=2in]{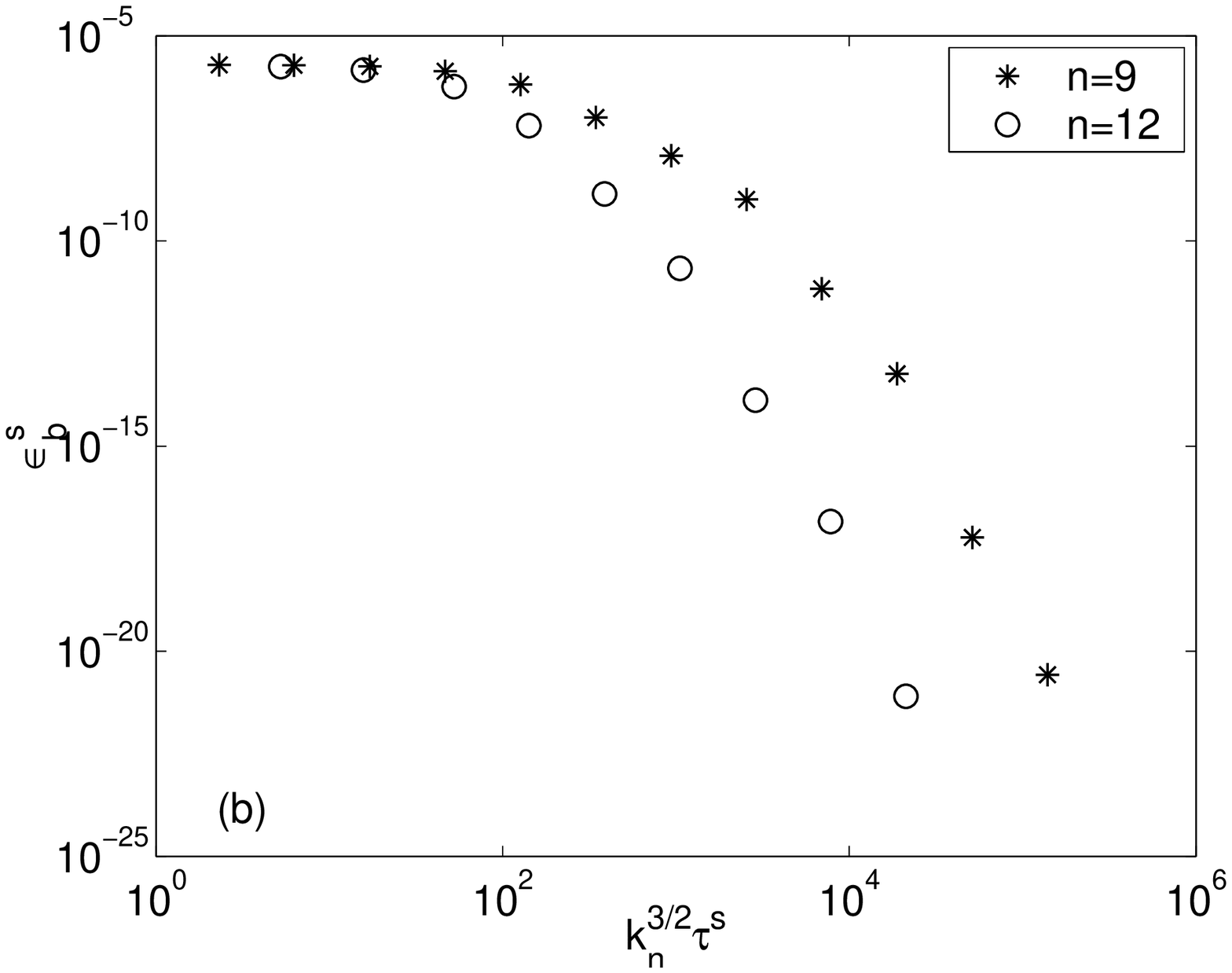}
\caption{\label{dc}(a) Plot of ${\mathcal E_b^s}/k_n$ versus $k_n^2\tau^s$
for shells $n=9$ and $n=12$ with $q=1$, illustrating the
data-collapse implied by the scaling form of Eq. (\ref{main}).\\(b) The 
analog for $q=0$ illustrating the lack of data collapse.}
\end{figure}

For $q=1$, Eq.~(\ref{main}) has the form 
${\mathcal E_a}^s(k_n,\tau^s)=k_n\hspace{0.2cm}\Phi_a(k^{2}_n\tau^s,1/
\nu^2,1/\eta^2)$. In FIG. \ref{dc}(a) we show on a log-log plot 
that a data-collapse occurs for representative shells, here 
$n=9$ and $n=12$, when ${\mathcal E_b}^s(k_n,\tau^s)/k_n$ is 
plotted against $k^2_n\tau^s$. In FIG. \ref{dc}(b) we show the analog for $q=0$ 
illustrating the lack of data collapse. Note, however, that the collapse 
improves as $k\rightarrow 0$; as we show below this leads to a power-law decay 
of $E_b^s$ over a limited range of $t$ and is related to the `permanence of 
large eddies'.

\begin{figure}
\includegraphics[height=2in]{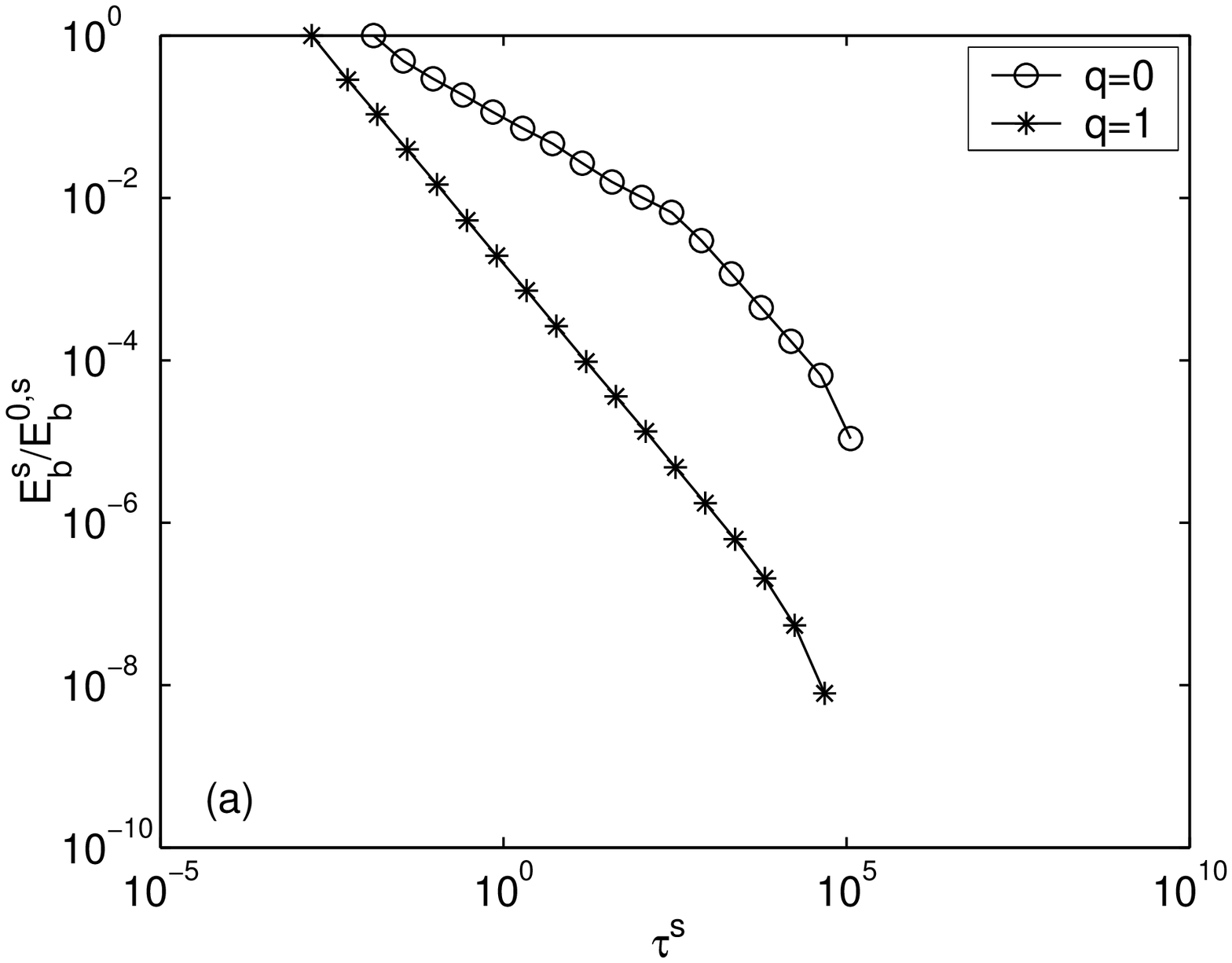}
\includegraphics[height=2in]{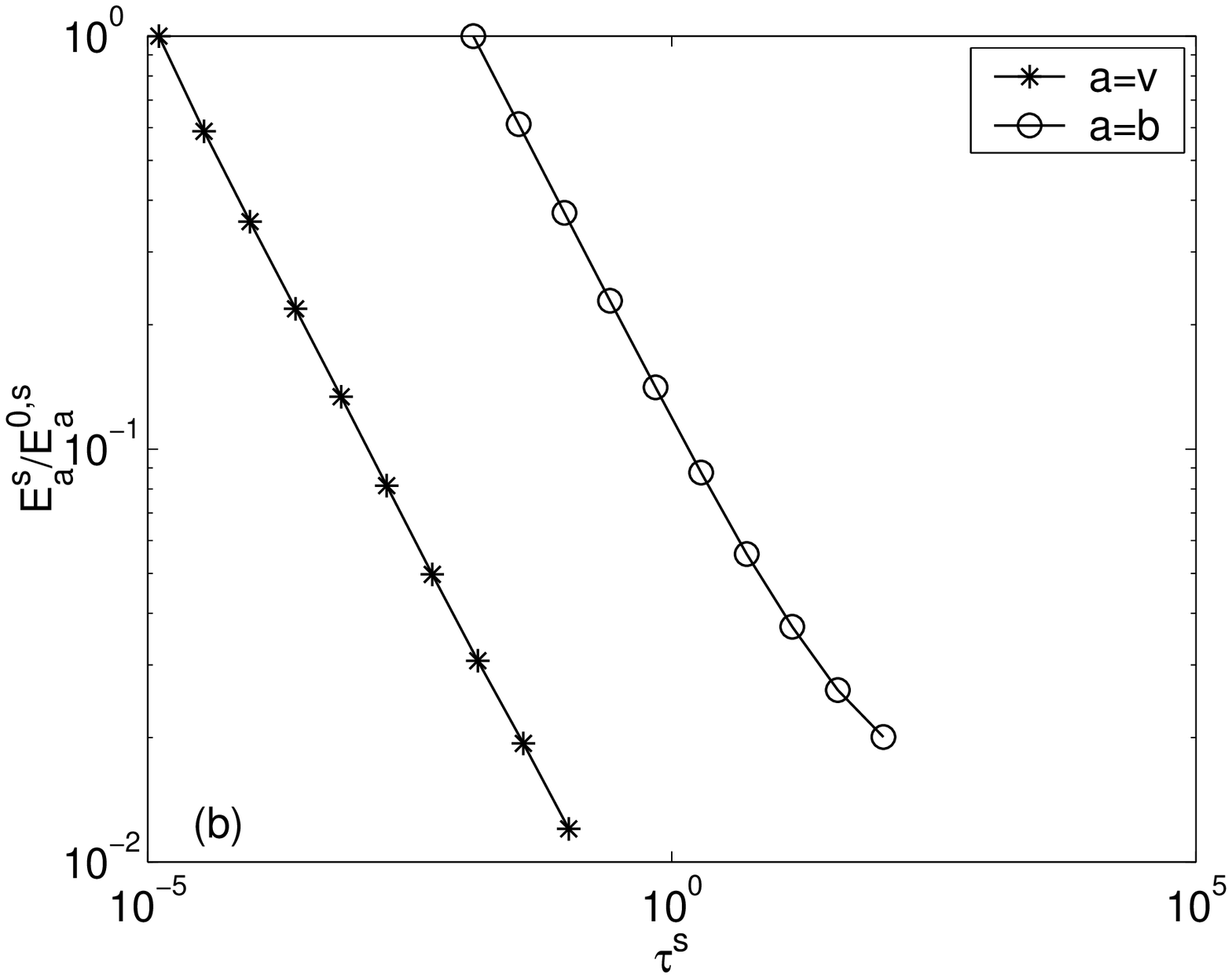}
\caption{\label{endec}(a) Log-log plots of the decay of the magnetic energy 
$E_b^s$ as a function of time $\tau^s$. The observed slope is 
$-1.000\pm0.001$ for $q=1$ in agreement with Eq. (\ref{eqb}). A nominal slope 
of $-0.66\pm0.01$ can be fit to the curve for $q=0$ in the range 
$\tau^s=1.28\times10^{-2}$ and $\tau^s=7.70\times10^2$.
(b) Log-log plots of the decay of the kinetic and 
magnetic energies ($E_v^s$ and $E_b^s$ respectively), without initial 
equipartition of energy ($E_v^{0,s}=10^3E_b^{0,s}$) for $q=0$.} 
\end{figure}

In FIG. \ref{endec}(a) we show the decay (for $q=0$ and $1$) of the magnetic 
energy 
$E_b^s$ with an observed slope of $-1.000\pm0.001$ (with errors from 
least-square fits) for $q=1$. The final point on this graph for $q=1$ 
exhibits the beginnings of the crossover at $t=t_*$ where $L_a^s(t)$ 
becomes comparable to $k_1^{-1}$, the analog of the system-size in the shell 
model. 

For $q=0$, a nominal slope of $-0.66\pm0.01$ may be fitted over a 
portion of the curve between $\tau^s=1.28\times10^{-2}$ and 
$\tau^s=7.70\times10^2$. 
This slope is in agreement with the law obtained from a hypothesis of 
`permanence of large eddies', wherein the viscous terms are negligible at 
small $k$ in Eq. (\ref{main}) leading to a decay 
law of $-0.667$. The hypothesis of `permanence of large eddies' implies that 
if $E_{\bf a}^0(k)\sim k^q$, for $k\rightarrow 0$, then the 
total energy decays as $E_T\sim(t-t_0)^{-2\left(\frac{1+q}{3+q}\right)}$ as 
discussed in the fluid turbulence context in Ref. \cite{Frisch}.

So far we have used initial conditions with initial equipartition, i.e., 
$E_v^{0,s}=E_b^{0,s}$. We find that this is maintained during the decay which 
is why we have shown plots of only the magnetic energy. For $q=1$ and 
$E_v^{0,s}\neq E_b^{0,s}$, both $E_v^s$ and $E_b^s$ decay as $t^{-1}$ for 
$0<t<t_*$ and maintain their initial ratio. In FIG.~\ref{endec}(b) 
we plot the temporal evolution of the kinetic and 
magnetic energies without initial equipartition, by setting 
$E_v^{0,s}=10^3E_b^{0,s}$ (here $Re_v^{0,s}=10^3Re_b^{0,s}$) for $q=0$. Unlike 
the case $q=1$, the kinetic and magnetic parts exchange energy here, but the 
dominant one (the kinetic energy in our plot) decays with a slope 
$-0.490\pm0.002$. 

\begin{figure}
\includegraphics[height=2in]{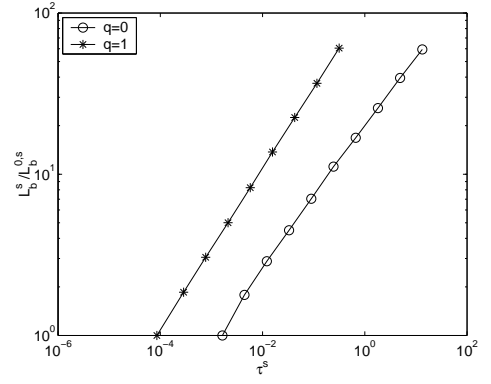}
\caption{\label{lambda} Log-log plots of the growth of the magnetic integral 
scale $L_b^s$ (for $q=0$ and $1$) as a function of time with 
$E_v^{0,s}=E_b^{0,s}$. The observed slope is $0.500\pm 0.002$ for $q=1$ in 
agreement with the magnetic integral-scale analog of Eq. (\ref{taylor}).}
\end{figure}

\begin{figure}
\includegraphics[height=2in]{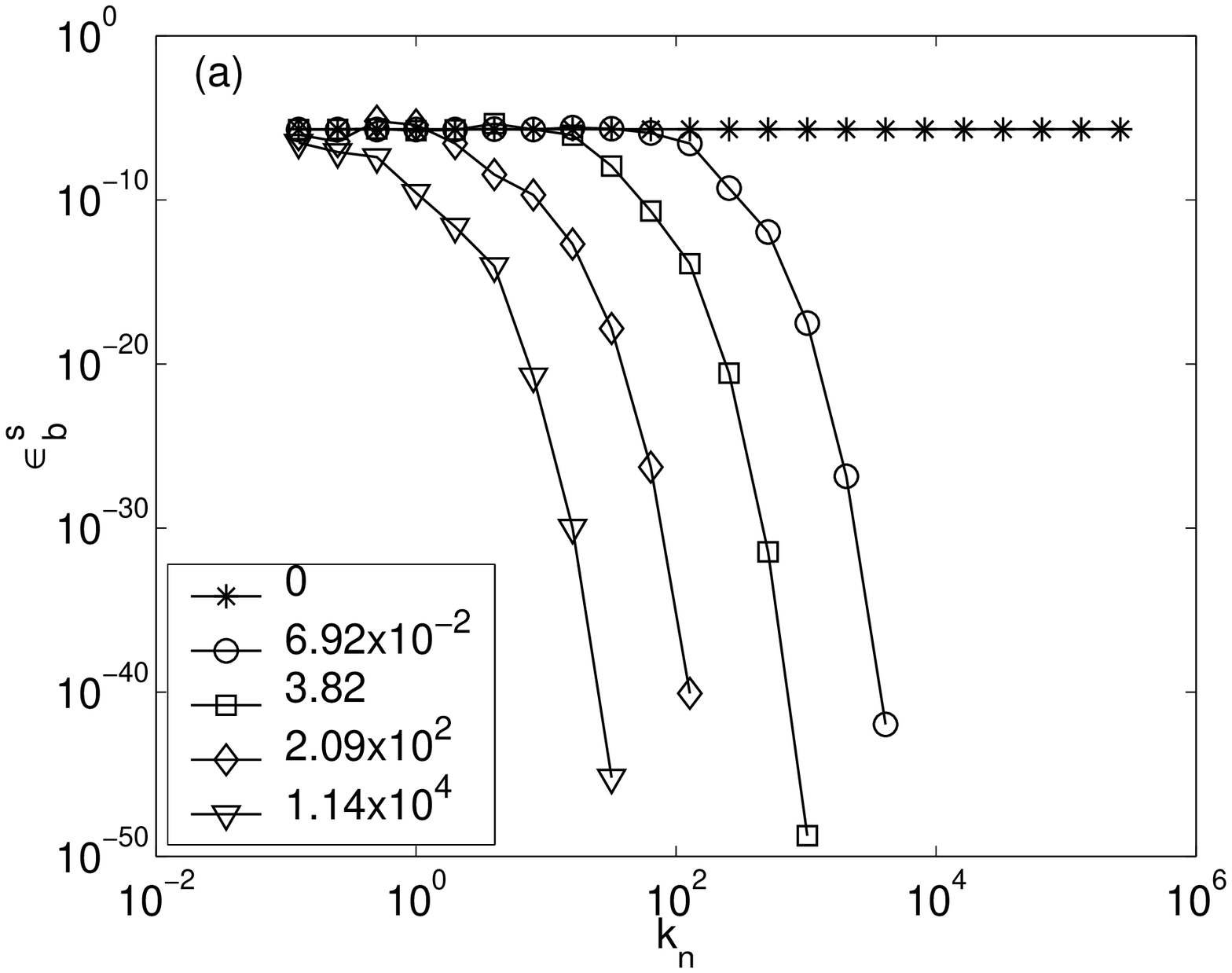}
\includegraphics[height=2in]{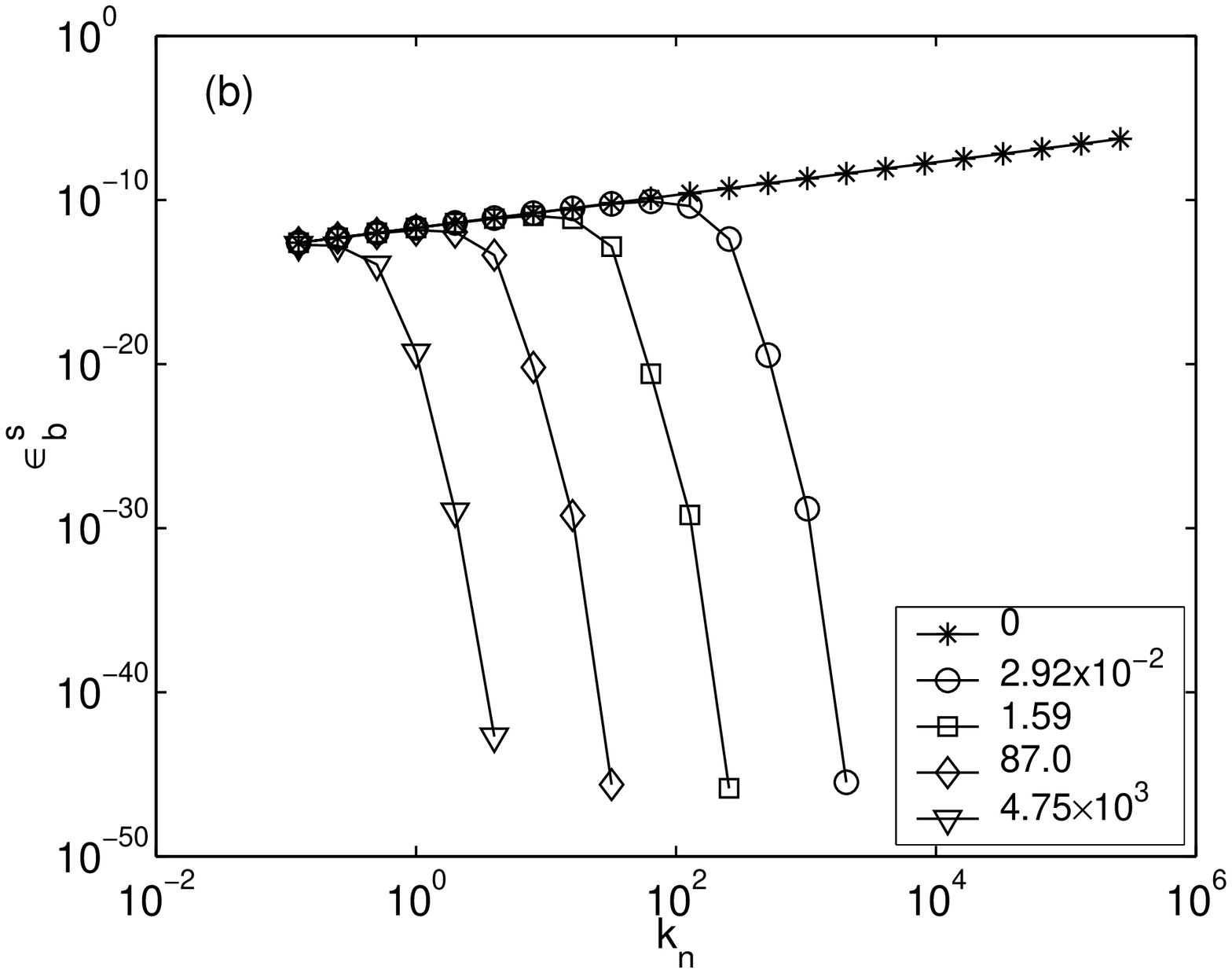}
\caption{\label{spectrashell} Log-log plots of the MHD shell model magnetic 
energy 
density ${\cal E}_b^s$ as a function of wavenumber (temporal sequence with 
$\tau^s$ values indicated in the legend) for (a) $q=0$ and (b) $q=1$.}
\end{figure}

In FIG.~\ref{lambda}, we plot (for $q=0$ and $1$) the shell model analog of 
the magnetic integral scale $L_b^s\equiv[\langle(\sum_n|b_n|^2/k_n)/
\sum_n|b_n|^2\rangle]$ as a function of time. The observed slope is 
$0.500\pm 0.002$ for $q=1$.
In FIGs. ~\ref{spectrashell}(a) and (b), we plot a temporal sequence 
(with seperation in units of $\tau^s$ indicated in the legend) of the 
magnetic-energy density ${\cal E}_b^s$ as a function of wavenumber for $q=0$ 
and $1$. The kinetic-energy density evolves in a similar manner for the case of 
 initial equipartition. 
Note that the evolution of these energy spectra illustrate quantitatively, the 
qualitative notion of `permanence of large eddies': As time progresses, the 
large-$k$ part of the spectra get modified by viscous effects; however, the 
power-law part at small $k$ maintains its initial form for $t\lesssim t_*$. We 
have also explicitly checked that the bounds (\ref{hc}) and (\ref{hm}) are 
respected in our shell-model simulations. 

For our DNS study of the 3DMHD equations, we use a pseudospectral 
method\cite{Dhar} to solve Eqs. (\ref{mhd}) in a cubical box 
of side $2\pi$ with periodic boundary conditions and $80^3$ Fourier modes. 
For the temporal evolution, we use an Adams-Bashforth scheme (step size 
$\delta t=0.02$) and $\nu=\eta=10^{-2}$ 
(we exclude any hyperviscosities). We define the initial fluid and magnetic 
Reynolds numbers 
to be $Re_{\bf v}^0\equiv2\pi v_{rms}^0/\nu$, $Re_{\bf b}^0
\equiv2\pi b_{rms}^0/\eta$ (here $Re_{\bf v}^0$ and $Re_{\bf b}^0$ equal 
$34$ for $q=1$ and $17$ for $q=0$) with 
$a^0_{rms}\equiv[\langle\sum_{{\bf k}}|{\bf a}
({\bf k})|^2\rangle]^{1/2}$ the root-mean-squares of the 
velocities and magnetic fields and initial large-eddy turnover times to be 
$\tau_{\bf a}\equiv2\pi/a_{rms}^0$. As in the shell-model case, $\tau_
{\bf v}=\tau_{\bf b}$ (here they equal $115.4$ for $q=1$ and $228.2$ for $q=0$) 
and $\tau\equiv t/\tau_{\bf a}$ ($t$ is the product of the number of iterations 
and $\delta t$). The initial velocity and magnetic fields are taken 
to be ${\bf v}({\bf k})\sim k^{q/2}e^{i\theta_{\bf k}}$, ${\bf b}({\bf k})
\sim k^{q/2}e^{i\phi_{\bf k}}$ with $\theta_{\bf k}$ and $\phi_{\bf k}$ 
being random variables distributed uniformly between $0$ and $2\pi$. Our DNS  
results are restricted to initial equipartition of energy between the velocity 
and the magnetic field. 

\begin{figure}
\includegraphics[height=2in]{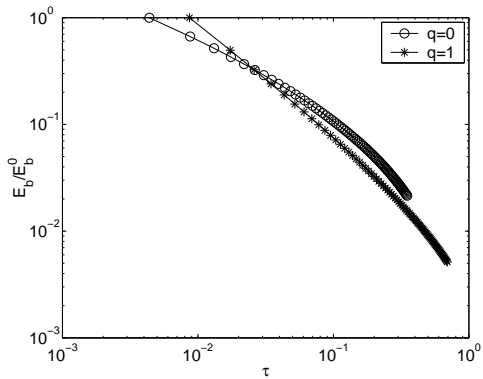}
\caption{\label{edecdns} Log-log plots of the magnetic energy $E_{\bf b}$ as a 
function of time for $q=0$ and $1$ for 3DMHD from our $80^3$ DNS.}
\end{figure}

\begin{figure}
\includegraphics[height=2in]{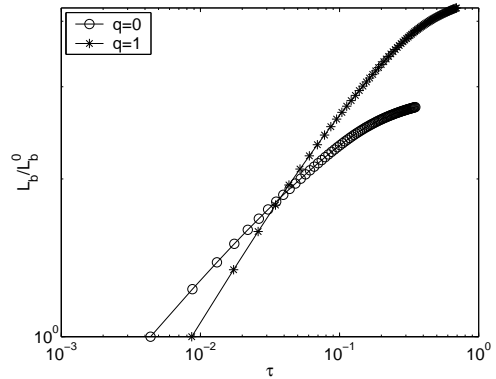}
\caption{\label{integraldns} Log-log plots of the magnetic integral scale 
$L_{\bf b}$ as a function of time for $q=0$ and $1$ for 3DMHD from our 
$80^3$ DNS.}
\end{figure}

In FIG. \ref{edecdns} we plot the magnetic energy $E_{\bf b}$ as a function of 
time for $q=0$ and $1$. A slope of $-1.057\pm 0.004$ is observed for 
$q=1$ between 
$\tau=8.7\times10^{-3}$ and $\tau=8.7\times10^{-2}$ in reasonable agreement 
with Eq. (\ref{eqb}). 
In FIG. \ref{integraldns} we plot the magnetic integral scale $L_{\bf b}$ as a 
function of time for $q=0$ and $1$. A slope of $0.401\pm 0.003$ is observed 
for $q=1$ between $\tau=8.7\times10^{-3}$ and $\tau=8.7\times10^{-2}$. We 
believe the slight discrepancy with the magnetic integral-scale analog of 
Eq. (\ref{taylor}) arises because of the limited spatial resolution of the 
DNS. 
\begin{figure}
\includegraphics[height=2in]{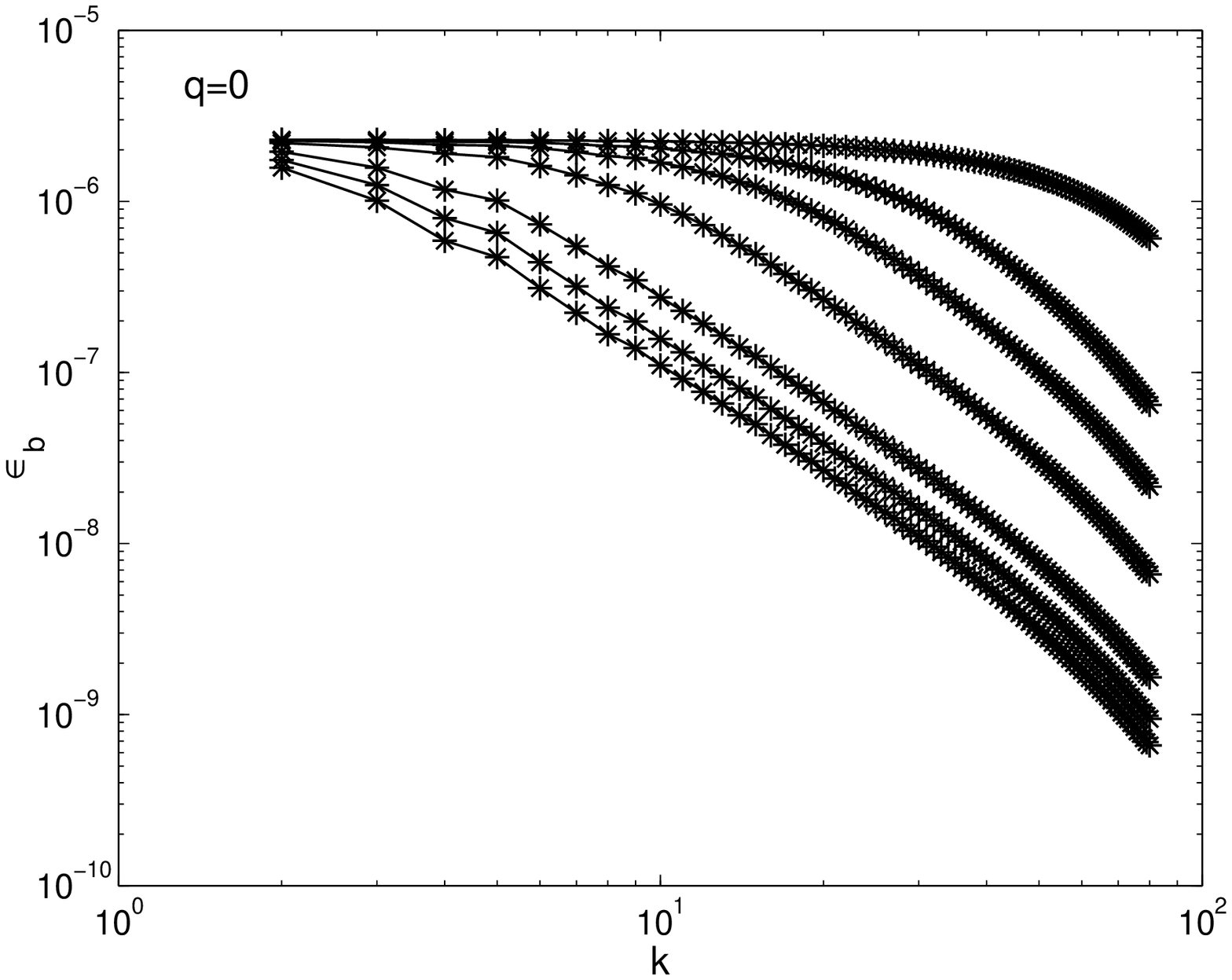}
\includegraphics[height=2in]{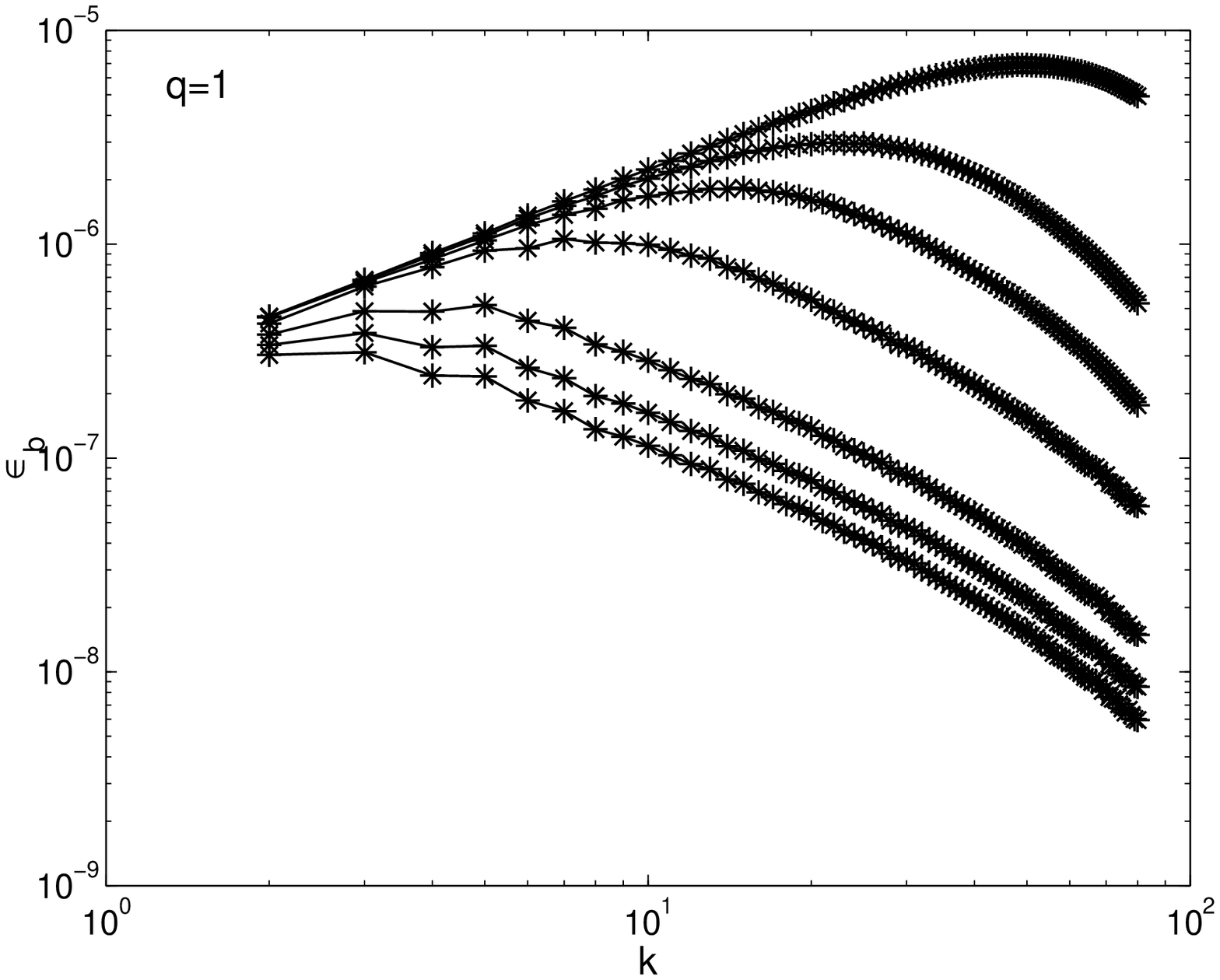}
\caption{\label{specdns} Log-log plots of the magnetic energy density ${\cal E}_
{\bf b}$ as a function of wavenumber $k$ (temporal sequence in steps of 
$0.02\tau$ for $q=0$ and $0.05\tau$ for $q=1$) for 3DMHD from our $80^3$ DNS.} 
\end{figure}
In FIGs. \ref{specdns}(a) and (b), we plot a temporal sequence 
(in steps of $0.02\tau$ ($q=0$) and $0.05\tau$ ($q=1$)) of the 
magnetic energy 
density ${\cal E}_{\bf b}$ as a function of wavenumber $k$ for $q=0$ and $1$. 
The small-$k$ part of the energy spectra in FIGs. \ref{specdns}(a) and (b) 
change somewhat more than their shell-model counterparts in FIG. 
\ref{spectrashell}. We believe this is because of the low spatial resolution 
of our DNS for 3DMHD. Our DNS results are presented here to complement our 
shell-model results and to show, in particular, that our general conclusions 
are not shell-model artifacts.
\section{Discussion and Conclusion}
To summarize, we have derived relations for the decay of the kinetic and 
magnetic energy, the growth of the Taylor and integral scales, and bounds for 
the cross- and magnetic helicity in unforced, incompressible, homogeneous and 
isotropic 3DMHD turbulence. We have confirmed our results numerically via 
shell-model and DNS studies. We show explicitly that our results about the 
power-law decay of the energies hold for times $t<t_*$, where $t_*$ is the 
time at which the integral scales become comparable to the system size. For 
$t<t_*$, our numerical results are consistent with those predicted by the 
principle of `permanence of large eddies'. Our study also has some 
implications for suggestions of `strong universality' in decaying turbulence 
as we discuss below.

In an early study\cite{Heisenberg}, an expression for the kinetic- 
energy density was hypothesized by assuming the existence of an 
`eddy-viscosity'. An energy decay similar to what we get for $q=1$ was then 
considered in the context of decaying fluid turbulence. However, this study
did not consider the initial-condition dependence of the decay laws or the 
effect of the growth of the relevant length scales.

In a recent study\cite{Lvov}, arguments have been given for `strong 
universality' in forced and decaying fluid turbulence in a shell model. 
For decaying turbulence, 
this means that scaling exponents of the $n$-th order velocity structure 
functions and their coefficients in the isotropic sector, normalised by the 
mean energy flux, are universal and the same as those for structure functions 
that are obtained for forced turbulence. For 
the purpose of our discussion here, this would imply that, 
irrespective of the initial condition used, after an initial period of decay, 
the energy spectrum should evolve towards one with a part that goes as 
$k^{-5/3}$ (aside from multiscaling corrections which we are not concerned 
with here). 
We have been able to get results similar to those of Ref. \cite{Lvov} by using 
the MHD shell model (Eqs. (\ref{shell})) and initial conditions as in Ref. 
\cite{Lvov}. In particular, if we start with equal initial kinetic and 
magnetic energies in the first two 
shells, the MHD-shell-model simulations exhibit an energy cascade to higher 
shells. Once this cascade process is complete, the energy densities display a 
part that can be fitted to a $k^{-5/3}$ form and the integral scale moves 
towards small wavenumbers. The point we would like to 
highlight here is that the evolution to a $k^{-5/3}$-type energy spectrum 
does not occur for the power-law initial conditions we have concentrated on in 
this paper. This is especially clear in the MHD-shell-model studies shown in 
FIG. \ref{spectrashell}. The reason for this is the `permanence of large 
eddies': For small $k$, the energy spectrum retains the power-law 
dependence of the initial energy spectrum. Deviations from this initial power 
law become significant only for times $t>t_*$, at which the time-dependent 
integral scale $L_{\bf a}(t)$ becomes comparable to the size of the system. 
Given the parameters, such as $\nu$ and $\eta$, that we have used, by 
time $t_*$, the Reynolds numbers is sufficiently low that a fresh 
energy-cascade to large 
$k$ is not established again. Thus, suggestions of `strong universality' should 
be made with caution, for they apply only to a class of initial conditions that 
does not include the power-law initial conditions used here for the decay of 
3DMHD turbulence.

The initial conditions used in Ref. \cite{Biskamp} are also qualitatively 
different from those that we use here in so far as they begin with a large 
energy in a few Fourier modes. Thus they develop energy cascades and do not 
explore the types of decay that are associated with the `permanence of large 
eddies' we concentrate on here.
\begin{acknowledgments}
C.K. thanks Tejas Kalelkar for useful discussions, CSIR (India) for 
financial support, SERC (IISc) for computational resources, and R.P., the 
Indo-French Centre for Promotion of Advanced Scientific Research 
(IFCPAR Project No. 2402-2) for support. We thank A. Basu and D. Mitra for 
discussions. 
\end{acknowledgments}


\begin{thebibliography}{11}
\bibitem{Isotropic} We restrict ourselves to statistical invariance under 
rotations but not necessarily under reflections.
\bibitem{Convergence} The energy densities decay exponentially at large 
wavenumbers in the far-dissipation range. We have therefore assumed 
convergence of the integrals appearing in Eqs. (11)-(14), which appears 
reasonable considering the presence of a naturally occuring viscous cutoff at 
large wavenumbers.
\bibitem{Erratum} There is a typographical error in the equations given in 
\cite{Basu}. This does not affect their results. We have corrected this error 
in Eq.~\ref{shell}.
\end{thebibliography}
\end{document}